# Screening Breakdown on the Route toward the Metal-Insulator Transition in Modulation Doped Si/SiGe Quantum Wells

Z. Wilamowski[1,2], N. Sandersfeld[1], W. Jantsch[1], D.Többen[3], F. Schäffler[1]

[1]*Institut für Halbleiterphysik, Johannes Kepler Universität, A-4040 Linz, Austria*

[2]*Institute of Physics, Polish Academy of Sciences, Al Lotnikow 32/46, PL 0668 Warsaw, Poland*

[3]*Walter-Schottky-Institut, Technische Universität München, D-85748 Garching, Germany*

Exploiting the spin resonance of two-dimensional (2D) electrons in SiGe/Si quantum wells we determine the carrier-density-dependence of the magnetic susceptibility $\chi_m$. Assuming weak interaction we evaluate the density of states at the Fermi level, $D(E_F)$, and the screening wave vector, $q_{TF}$. Both are constant at higher carrier densities $n_s$, as for an ideal 2D carrier gas. For $n_s < 3 \cdot 10^{11}$ cm$^{-2}$, they decrease and extrapolate to zero at $n_s = 7 \cdot 10^{10}$ cm$^{-2}$. Calculating the mobility from $q_{TF}$ yields good agreement with experimental values justifying the approach. The decrease in $D(E_F)$ is explained by potential fluctuations which lead to tail states that make screening less efficient and - in a positive feedback - cause an increase of the potential fluctuations. Even in our high mobility samples the fluctuations exceed the electron-electron (e$^-$-e$^-$) interaction leading to the formation of puddles of mobile carriers with at least 1 µm diameter.

Experimental evidence for a low-temperature metallic state in a variety of 2D carrier systems[1,2,3,4,5,6,7] (2DCS) have newly highlighted the long-lasting discussions about the nature of the metal-to-insulator transition (MIT). Since scaling theory for non-interacting carrier systems predicts localization in 2D at T=0,[8] recent experimental and theoretical efforts concentrate on interacting 2DCS´s. For these the carrier density $n_s$ is a most important parameter, because it determines the ratio $r_s$ between the e$^-$-e$^-$ Coulomb energy and the kinetic energy $E_{kin}$ as[9] $r_s = 1/(a_B\sqrt{\pi n_s})$, with the effective Bohr radius, $a_B$ =32 Å for Si. The other important quantity is disorder which we characterize here by the amplitude of potential fluctuations, $\delta V$ and the dimensionless ratio, $r_d = e\delta V/E_{kin}$, normalized again to $E_{kin}$. (Here e is the electron charge). If disorder is negligible ($r_d \ll r_s$) collective localization (Wigner crystallization) is predicted[6,9] at around $r_s$=37, whereas increasing $r_d$ leads to carrier localization at lower $r_s$ via the defect potential. Most of the experiments reported[6] employed disordered systems with $r_s$ values around 10 at the MIT. Moreover, the experiments almost exclusively concentrated on transport, which allows the identification of an MIT, but does not reveal the mechanisms that cause it. It is therefore not surprising that the origin of the apparent metallic phase is heavily debated.[10,11,12]

To gain additional experimental information, we propose a new method that gives access to $r_d$ and the screening properties on the metallic side of the MIT. It is based on conduction electron spin resonance (CESR) in strained Si quantum wells, and on the magnetic susceptibility $\chi_m$ of the 2DCS determined from CESR. Far enough on the metallic side a weakly interacting 2DCS can be assumed in a first approximation. In this limit, and at low temperatures, the 2D carrier gas exhibits Pauli paramagnetism, and thus $\chi_m$, is proportional to the density of states (DOS) at the Fermi level, $D(E_F)$:[13]

$$\chi_m = g^2\mu_B^2 \cdot D(E_F) \quad \text{(for } kT \ll E_F\text{),} \quad (1)$$

where g is the Landé factor and $\mu_B$ Bohr's magneton. In the same limit, the Thomas-Fermi screening wave vector $q_{TF}$ reads:[14]

$$q_{TF} = \frac{e^2}{2\varepsilon\varepsilon_0} D(E_F) \quad \text{(for } kT \ll E_F\text{)}, \quad (2)$$

with the permittivity $\varepsilon\varepsilon_0$ of the medium. In this limit, obviously, $\chi_m$ is directly proportional to $q_{TF}$ and thus CESR allows for an experimental evaluation of $q_{TF}$. Although this simple model is not expected to hold all the way to the MIT, we show that the carrier mobilities calculated from the experimental $q_{TF}$ agree over a large $r_s$ range surprisingly well with the measured mobilities justifying this approach *a posteriori*.

The experiments exploit the recently identified CESR signal of a 2D electron gas in the strained Si channel of a modulation-doped SiGe/Si/SiGe double heterostructure.[15,16,17] The 2DCS in this heterostructure has the same twofold valley degeneracy and effective transport mass as the inversion layers of the more widely studied MOSFETs,[1,7] but modulation doping allows for much higher mobilities.[18] Other important differences comprise the lack of oxide charges,[10] the reduced interface roughness, and the presumably much lower density of interface charges at the crystalline Si/SiGe interfaces. The negligible spin-orbit interaction, and the lack of magnetic impurities lead to extremely long spin relaxation times of several µs,[19,20] which provide exceptionally narrow ESR signals, and thus an unusually low detection limit of $3\cdot 10^9$ spins/cm$^2$.

The ESR experiments were performed at 4.2 K in a standard X-band spectrometer (Bruker 200 SRC) with a TE102 cavity. The samples were grown by molecular beam epitaxy on 1000 Ωcm Si(001) substrates, which show complete carrier freeze-out at 4.2 K. A 20 nm thick Si channel with tensile in-plane strain was deposited on a strain-relaxed $Si_{0.75}Ge_{0.25}$ buffer layer, which consists of a 0.5 µm thick $Si_{0.75}Ge_{0.25}$ layer on top of a 2 µm thick $Si_{1-x}Ge_x$ layer with compositional grading.[16,18] The upper $Si_{0.75}Ge_{0.25}$ barrier was modulation doped with a 12.5 nm thick, nominally undoped spacer layer, and capped with 5 nm of Si. Sample 1 has a low Sb doping concentration of $5\cdot 10^{17}$ cm$^{-3}$ in a 30 nm wide supply layer. It shows strong persistent photoconductivity: After cooling in darkness to 4.2 K the carrier concentration $n_s$ is less than $1\cdot 10^{11}$ cm$^{-2}$, but it increases with the dose of Si band gap illumination, and saturates at $n_s = 3\cdot 10^{11}$ cm$^{-2}$. Sample 2 from the same wafer had a palladium Schottky gate, and sample 3 was higher doped, leading to a constant $n_s = 7\cdot 10^{11}$ cm$^{-2}$. Sample 3 has a mobility of about 200,000 cm$^2$/Vs at 4.2 K, whereas the mobilities of the other samples are strongly $n_s$ dependent (see Fig. 2 below).

All spectra were taken with the magnetic field perpendicular to the 2DCS which in this geometry yields a broad cyclotron resonance (CR) absorption and a very narrow CESR signal[16,20] (see insert in Fig. 1). In sample 1, $n_s$ was adjusted by the illumination dose, and monitored by fitting the CR signal shape.[19] In the gated sample 2, $n_s$ was determined by gate-voltage-dependent Shubnikov-de-Haas experiments.

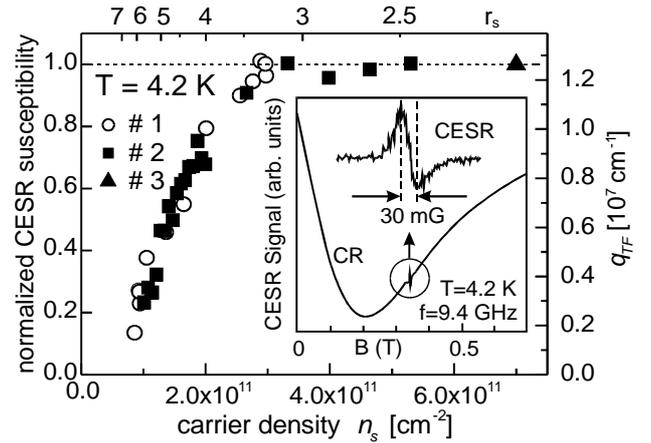

Fig.1: Integral CESR absorption normalized to the saturation value that corresponds to the Pauli susceptibility of an ideal 2DCS. The right ordinate has been converted into the Thomas-Fermi wave vector $q_{TF}$. The insert shows a typical spectrum with B perpendicular to the 2DCS; CR labels the cyclotron resonance signal. The top scale gives $r_s$ (see text).

Fig. 1 shows the integral CESR absorption, and thus $\chi_m$, normalized to its saturation value, which is the same for all samples, as a function of $n_s$. Good agreement is found between the gated and illumination-dosed samples. Following Eqs (1) and (2), $\chi_m$ can be converted into an absolute value of $q_{TF}$, as given by the right hand side ordinate in Fig 1. The constant of proportionality was determined in the saturation range by assigning to it the constant value of the ideal 2DCS: $q_{TF}^{2D} = \frac{2g_v}{a_B} = 1.27\cdot 10^7$ cm$^{-1}$, with $g_v = 2$ in tensely strained Si. The striking feature in Fig. 1 is that $q_{TF}$ reaches the intrinsic value of an ideal 2DCS only for $n_s > 3\cdot 10^{11}$ cm$^{-2}$, whereas it decreases monotonously below that concentration and extrapolates to zero at a critical density of $n_s^c \cong 7\cdot 10^{10}$ cm$^{-2}$. This compares well with the critical MIT carrier concentration of $9\cdot 10^{10}$ cm$^{-2}$ reported for high-mobility Si MOSFETs.[1]

The Thomas-Fermi approximation may become inadequate for describing transport close to the MIT, where $r_s$ is 7 in our samples. It is therefore important to assess the $n_s$ range over which $q_{TF}$ derived from the CESR experiments is relevant for the transport properties. For that purpose we measured the Hall mobility of gated and illumination-dosed samples as a function of $n_s$ (Fig. 2, solid squares). Below $4.5\cdot 10^{11}$ cm$^{-2}$ the experimental mobility follows a power law in $n_s$ with a fitted exponent of 2.34 (solid line). For comparison, we calculated the Thomas-Fermi transport scattering time[14] replacing the constant $q_{TF}^{2D}$ of the ideal 2D electron gas with the experimentally determined $q_{TF}(n_s)$ from Fig. 1. For simplicity a Howard-Fang envelope function for the wave function perpendicular to the well was assumed.[14] The resulting mobilities are plotted as open symbols in Fig. 2.

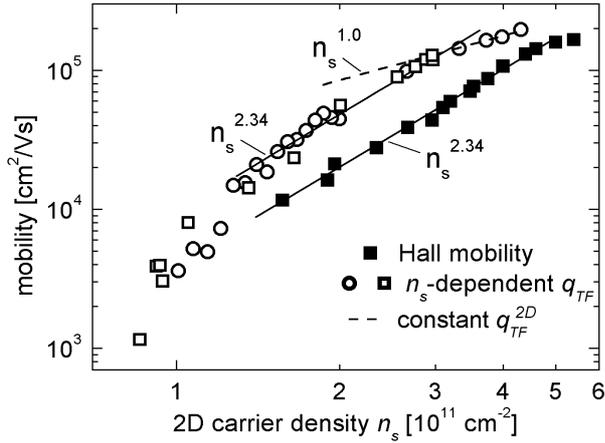

Fig. 2: Hall mobility (solid symbols) and calculated mobility in Thomas Fermi approximation (open symbols) versus $n_s$. The dashed line represents the ideal 2DCS. Solid lines are fits to the data and represent an $n_s^{2.34}$ power law.

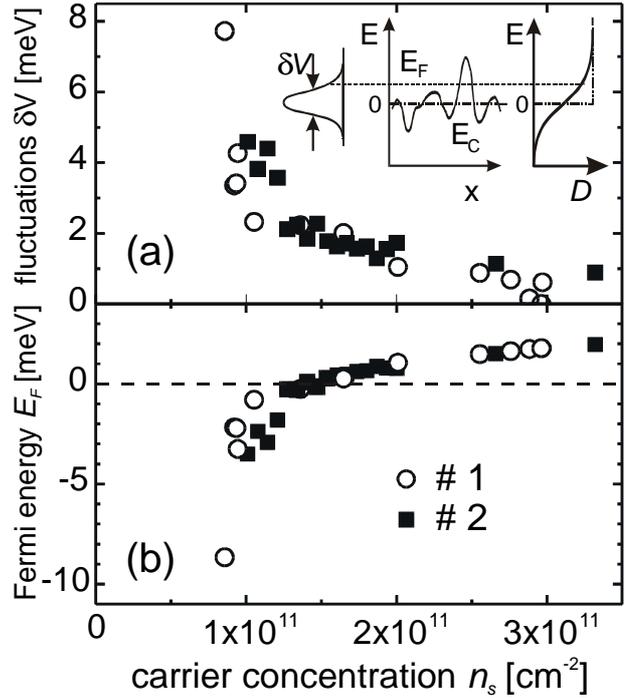

Fig. 3: Fluctuation amplitude $\delta V$ (a), and Fermi energy $E_F$ (b), as a function of the carrier density $n_s$ for samples 1 and 2. The inset shows the Gaussian distribution of conduction band edge fluctuations ($E_C$) with respect to $E_F$ (dotted), and the resulting DOS (solid) in comparison to the ideal DOS (dash-dotted).

The absolute values of the calculated and experimental mobilities differ by almost a factor of two, but this is not too surprising, since only remote impurity scattering has been considered. Much more striking is the fact that the Thomas-Fermi mobility perfectly reproduces the experimentally found $n_s^{2.34}$ power law down to about $1.5 \cdot 10^{11}$ cm$^{-2}$. For $n_s > 3 \cdot 10^{11}$ cm$^{-2}$, the calculated mobility bends over to the $n_s^{1.0}$ relation of an ideal 2DCS with constant $q_{TF}(E_F)$ (dashed line). The excellent agreement between the fitted power laws suggests that the $n_s$-dependent $q_{TF}$ determined from CESR is meaningful for transport experiments down to about $1.5 \cdot 10^{11}$ cm$^{-2}$.

In order to explain the observed decrease of $\chi_m$ and the associated breakdown in screening efficiency ($\chi_m \propto q_{TF}(E_F)$) as $n_s \rightarrow n_s^c$ we consider the effect of disorder. The latter introduces a low energy tail to the ideally sharp onset of the DOS which we model by a Gaussian distribution of the potential fluctuations that are superimposed on the band edge. The width of the distribution corresponds to the screened fluctuation amplitude $\delta V$. Since $D(E_F) \propto \chi_m$ is directly measured by the CESR signal according to Eq. 1, the two parameters $\delta V$ and $E_F$ can be determined as an unambiguous function of $n_s$ by simultaneously fitting the experimental data for $\chi_m$ and $n_s$ applying Fermi statistics (see insert of Fig. 3).[19]

Figs. 3a and 3b show results of $\delta V$ and $E_F$ versus $n_s$, with $E_F$ being measured relative to the band edge of the ideal 2D electron gas. For higher $n_s$, where screening is effective, $e\delta V$ remains smaller than $E_F$ ($r_d < 1$). At lower $n_s$, $\delta V$ increases drastically and finally it diverges. Simultaneously, $E_F$ becomes negative at about $1.5 \cdot 10^{11}$ cm$^{-2}$, i.e., it drops below the ideal band edge and thus enters the tail states. These results show that the drop of $E_F$, which was also observed in other experiments near the MIT,[21,22] is caused by the increase of $\delta V$ and the concomitant formation of tail states.

Close to $n_s^c$ the calculated mobilities in Fig. 2 drop exponentially. We attribute this behavior to a positive feed-back mechanism[11] that is driven by the loss of screening and the concomitant increase of the potential fluctuations. There $e\delta V$ exceeds $E_F$ (i.e., $r_d > 1$), which means that the 2DCS becomes disrupted. Finally, when $\delta V$ diverges, the system ceases to form a continuous 2DCS.

From the data presented so far it is not clear whether the diverging potential fluctuations at low $n_s$ are long or short ranged. In the first case the 2DCS would break apart into percolating 2D puddles, whereas the latter situation would lead to Anderson localization. The following observations indicate puddle formation. The CESR line width at the lowest carrier concentrations investigated is about 30 mG.[16,20] This implies a lower limit for the area over which carriers can move freely, because the minimum CESR line width is given by the hyperfine interaction with nuclear spins in the probing area. With the relative abundance of 4.7% of $^{29}$Si (the only stable Si isotope with nuclear spin) and the hyperfine constant of P in Si,[23] we estimate a minimum puddle size on the order of 1 μm$^2$. This value is in good agreement with spatially resolved compressibility measurements on GaAs hole channels.[22]

Rather large puddles are also consistent with transport experiments on similar samples[24] that revealed in the metallic regime Dingle ratios well in excess of 10. At such high values potential fluctuations are mainly due to the Coulomb potential of the ionized donors in the remote doping layer. These cause inherently smooth fluctuations, and thus large puddles in the insulating regime, since short-range fluctuations decay exponentially with increasing spacer thickness.[14] The mean e$^-$-e$^-$ spacing at $n_s$ = 7·10$^{10}$ cm$^{-2}$ is $\ell_s = 2/\sqrt{\pi n_s}$ = 430 Å, substantially longer than the donor-electron distance of the 125 Å (the thickness of the spacer layer). Therefore we expect from this simple argument already that the donor-electron interaction exceeds the e$^-$-e$^-$ interaction ($r_d \gg r_s$) as we see also in our results. Similar relations should apply to most of the reported material systems with $r_s$ values around 10.

In summary, by combining CESR and transport experiments we have studied high-mobility 2DCS´s in Si/SiGe heterostructures on their way from the metallic to the insulating state. We found that above $n_s$ = 1.5·10$^{11}$ cm$^{-2}$ the 2DCS is reasonably well described by Thomas-Fermi screening in the presence of weak potential fluctuations. At lower $n_s$ the potential fluctuations begin to diverge, and the 2DCS breaks apart. The experiments provide strong evidence for smooth potential variations that lead to the formation of 2DCS puddles rather than to Anderson localization. This picture is consistent with a recent model of Meir[25], who explained experimental transport results in the vicinity of the MIT by assuming percolating 2D carrier puddles linked by one-dimensional quantum point contacts. Thus, even in our high mobility modulation doped samples scattering is obviously dominated by potential fluctuations and the loss of screening rather than by the e$^-$-e$^-$ interaction.

**Acknowledgements:** Valuable discussions with G. Abstreiter, G. Brunthaler, A. Prinz and S.Lyon are gratefully acknowledged. Work supported by the *Fonds zur Förderung der Wissenschaftlichen Forschung, GMe*, and *ÖAD* (all Vienna), and by the KBN grant 8 T 11B 003 15 (Poland).